# INTERBLOCK REDUNDANCY REDUCTION USING QUADTREES

*Marcos Faúndez Zanuy, Xavi Domingo Reguant*
Department of Signal Theory and Communications (UPC)
c/Gran Capità S/N    (Campus Nord, Edifici D5)
08034 BARCELONA (SPAIN)
tel:+34 3 4016453 fax:+34 3 4016447
e-mail:marcos@gps.tsc.upc.es

**ABSTRACT**

This paper applies the quadtree structure for image coding. The goal is to adapt the block size and thus to increase the compression ratio (without reducing SNR). Also, the computational time is not significatively increased.

It has been applied to Block Truncation Coding of still images, and motion vector coding (interframe). An inter/intraframe application is also discussed.

## 1. INTRODUCTION

Usually, image compression algorithms are based on blocks. This alternative, presents two problems:
1. There is a compromise with the block size: compression factor vs quality.
2. It is not considered the redundancy between contiguous blocks.
These drawbacks can be resolved using adaptive block size: the block size is increased in homogeneous zones, and reduced in detailed zones. Normally, this is achieved increasing dramatically the computational complexity.

We propose a method based on block compression with small block size, and the clustering of blocks whenever they represent the same information. The clustering information is easily computed and coded, using quadtree structures. We have evaluated this algorithm with two coding methods:
a) Block truncation coding (BTC).
b) inter/intraframe coding with motion compensation.

This paper is organized as follows: In section two a still image coding application is presented. Section three deals an application for image sequence coding. Section four suggests other possibilities, and section five summarizes the main conclusions.

## 2. BTC + QUADTREES

We have applied quadtree structures to still images compressed with:
- basic BTC [1],
- a combination of IBTC2 [2][4] plus simultaneous σ-η compression [3][4] that we will name SNIBTC2.

Let us make a fast review of how BTC works. We divide the image in blocks of a convenient size (in our case 4x4 pels). We calculate the mean ($X$) and the 2$^{nd}$ order moment of the block grey levels ($X_2$). Then we code every block using a 1 bit/pel bit plane plus a low and a high grey levels (a and b). These three data are chosen in such a way that preserves the statistical information stated above. A bit plane bit is set to 1 when the grey level of the corresponding pel is greater than the mean of the block and to 0 when is lower than it. In the decoder, we restore the pels with bit plane value 1 to a and the other ones to b. The formulae to calculate a and b can be deducted knowing the bit plane, the mean and the 2$^{nd}$ order moment by imposing the condition of preserving statistics in the restored block. They can be computed with the following equations:

$$b = X - \sigma\sqrt{\frac{q}{n-q}} \quad ; \quad a = X + \sigma\sqrt{\frac{n-q}{q}}$$

$$\sigma = \sqrt{X^2 - X_2^2}$$

Note that $n$ is the number of bits in the bit plane and $q$ the number of them set to 1.

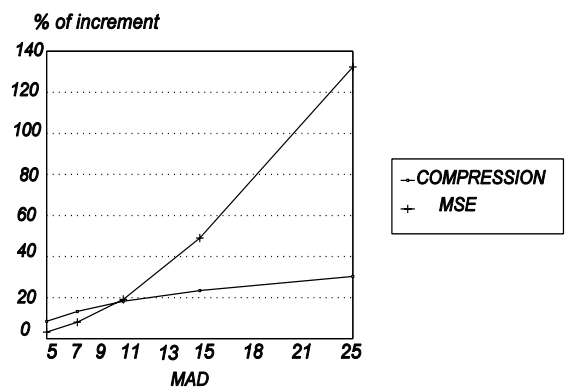

figure 1. Increment of compression & Mean Square error versus Maximum absolute difference threshold.

We have chosen the BTC algorithm because it is a very fast algorithm, and many variants have been proposed in order to increase compression, as basic BTC compression ratio is only 4:1 assuming 4x4 pels blocks. The algorithm that we propose is another way to increase compression without incrementing significatively the coding and decoding times.

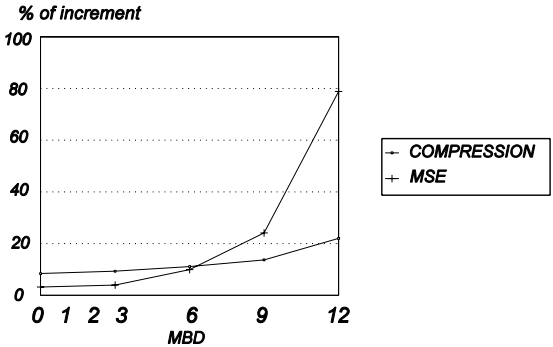

figure 2. Increment of compression & Mean Square Error versus Maximum Binary Distance Threshold.

In order to cluster the information of neighbouring blocks of the compressed BTC file we have used quadtrees with four levels. This implies that we may reduce the information of $4^{4-1}=64$ blocks to that of a single block whenever they all resemble each other.

As we explained before, each block is coded with a bit plane and two grey levels (a and b). Usually it is difficult to cluster jointly these three data, thus we have used three independent quadtrees to cluster them separately, in order to achieve higher compression. It is a too strict condition to allow clustering of the four neighbouring blocks only when they have identical information. So, we have defined relaxed conditions to cluster grey levels and bit planes. We will group grey levels if they differ in less than MAD (Maximum Absolute Difference), and bit planes if their binary distance is smaller than MBD(Maximum Binary Distance). Figure 1 and 2 show how compression and MSE increase with MAD and MBD, respectively. The percentages of increment are referred to the basic BTC compression ratio and MSE using image Lenna. Other images show similar results. Visual evaluation suggests to use MAD<8 to reduce false contouring effects in regions with low gradient gray scales. Increasing MBD produces visible errors in high contrasted blocks, where dark or light pixels appear where they must not.

Table 1 shows error and compression measures of the tested images using MAD=5 and MBD=6. Note that MSE(BTC) is the error of basic BTC compressed images before quadtrees application. The column labelled C shows the increase of compression in % with respect to the BTC compression ratio (4:1). We see how compression depends in a great manner on image details. However, with our implementation, coding and decoding the quadtrees takes only a 30% of the BTC compression time. We have used these definitions for error measures:

$$MSE = \frac{1}{n} \sum_{i=1}^{n} (x_i - \hat{x}_i)^2$$

$$MAE = \frac{1}{n} \sum_{i=1}^{n} |x_i - \hat{x}_i|$$

$$SNR(dB) = 20 \log \left( \frac{2^N - 1}{\sigma} \right)$$

$$\sigma = \sqrt{MSE - MAE} \approx \sqrt{MSE}$$

Figure 3 represents the distribution of blocks in the four levels of quadtree for images Lenna and Cheryl. Level 0 is the root level and represents the clustering of 64 blocks, while level three represents the non grouped blocks. Thus, bars indicate the number of blocks clustered in each level. Note that, as Cheryl has bigger uniform regions than Lenna, levels zero and one are often used, thus increasing compression.

Using quadtrees with more levels is neither necessary nor useful, because the overheads to code the trees grow in length and sometimes this leads to a reduction more than to an increase of compression.

We have developed SNIBTC2 [4] combining a double bit plane interpolation with a 10 bit simultaneous coding of the standard deviation and the mean of the blocks. This reduces the number of bits/block from 32 to 14, giving a constant 9.14:1 compression ratio. The error and time cost increases are little compared to the gain in compression. Note that using this technique the grey levels a and b to restore the blocks are computed in the decoder using the mean and the deviation restored from the compressed file.

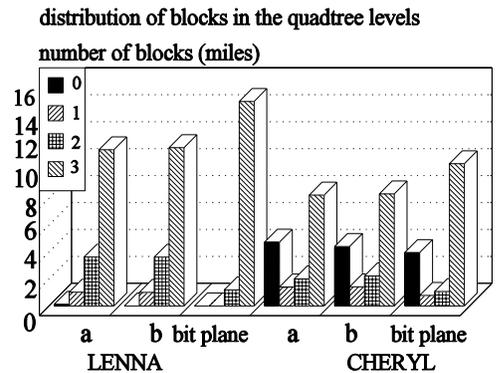

figure 3

We have used two independent quadtrees with SNIBTC2. One for the decimated bit planes and another for the 10 bit codes. The clustering criterion of this 10 bit codes consists in comparing separately the standard deviation and the first order moment. We group neighbour blocks information only when these two data are similar.

Applying quadtrees to the SNIBTC2 compressed files we obtain the results summarized in Table 2. Last row shows the results published in [2] using Vector Quantization with the grey levels and bit plane interpolation with BTC. Note that our results are very similar for the same image. However,

| IMAGE | MSE BTC | MSE | MAE | SNR (dB) | C (%) |
|---|---|---|---|---|---|
| Lenna | 44.5 | 48.9 | 4.2 | 31.2 | 11.1 |
| Pepper | 32.5 | 36.7 | 3.6 | 32.5 | 10.0 |
| Cheryl | 21.6 | 26.4 | 2.3 | 33.9 | 40.0 |

Table 1

| IMAGE | MSE | MAE | SNR (dB) | C | bpp |
|---|---|---|---|---|---|
| Lenna | 77.1 | 5.5 | 29.3 | 11.2 | 0.7 |
| Pepper | 67.0 | 5.0 | 29.9 | 11.0 | 0.7 |
| Cheryl | 43.2 | 2.8 | 31.8 | 12.7 | 0.6 |
| Lenna [2] | 76.4 | 5.5 | 29.3 | 10.7 | 0.8 |

table 2

quadtrees application is a very fast process, while VQ needs a time expensive codebook generation.

## 3. INTER/INTRAFRAME CODING.

In interframe coding applications motion compensation is usually utilized, and a great number of contiguous blocks present the same displacement vector. Thus, they can be clustered with a quadtree. In our experiments we have used a block size of 16x16 and 8x8 pixels, and images of 256x256 pixels. With smaller block sizes there is not a significative improvement [5] and the clustering of contiguous vectors is more critical. With higher block sizes the traslational motion model is not valid. Applying the quadtree to encode the displacement vector information, we achieve a typical compression factor 3:1 to 2:1 in videoconference images (considering only vector information and quadtrees versus vector information without quadtrees) with lossless coding.

Figure 5 shows a sample of the obtained results for the first two images of the secretary sequence. Similar results are obtained with other videoconference images. The left diagram uses three quadtree levels (from 16x16 to 64x64). It is interesting to observe that there are eight blocks of 64x64 which represent 128 blocks of 16x16 size, and the image is divided in 256 blocks of 16x16 pixels so half of the image is coded with a block size of 64x64. The right diagram uses four quadtree levels (from 8x8 to 64x64), and there is only one block of 64x64, which clusters 64 blocks of 8x8 pixels over a total number of 1024 blocks.

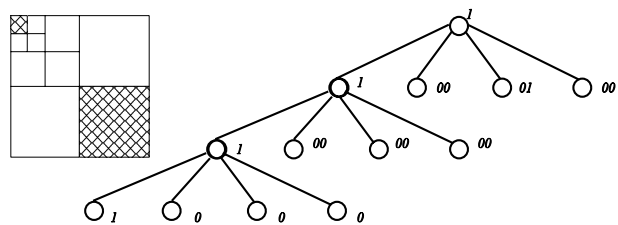

fig. 4 modified quadtree for inter/intraframe predictor. Last level and second bit of terminal nodes codifies the inter/intra frame decision. Dashed squares are intraframe coded zones.

Obviously if the clustering condition is "equal information" of clustered blocks, the error of the basic compression method is not modified. More interesting results are obtained if the tree is used to encode jointly the block shape and the inter/intraframe decision (see fig.4). For an inter/intraframe predictor, we propose the following structure in order to encode jointly the block shape and the inter/intraframe decision:
1. To implement the quadtree coding proposed in the last sections, grouping blocks coded with the same predictor.
2. Last level of the tree is used to indicate what kind of predictor has been used (inter or intra).

This last level was not coded in the original quadtree structure, because it shows no division, and this is already known from the number of levels used. Also, in the terminal nodes different of the last level, two bits are needed. One bit indicates terminal node, and the other the inter/intraframe decision. For instance, if a 0 represents a terminal node (no division), one alternative would be:
00: terminal interframe node
01: terminal intraframe node.

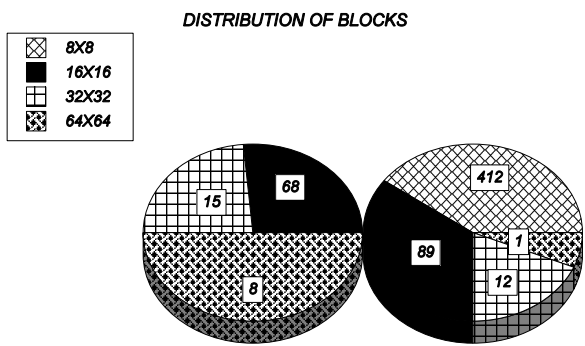

figure 5 Number of blocks of each size and portion of equivalent smaller blocks.

## 4. OTHER POSSIBILITIES.

These algorithms can be implemented with other tree structures, that let more flexibility in the block shape (for instance, rectangles), but more flexibility implies more bits for tree coding.

Other compression methods based on blocks, can utilize this structure to reduce the interblock redundancy, with a small adaptation.

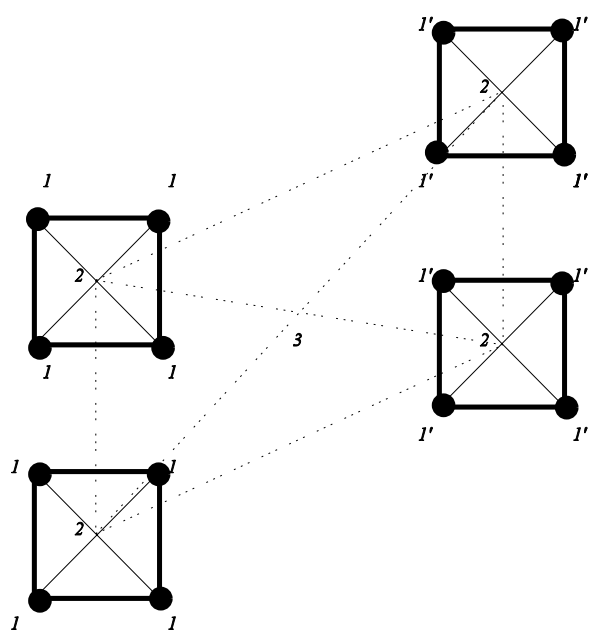

fig. 6  3D quadtree

The quadtree can be implemented in a 3D approach (see fig. 6). For interframe applications, displacement vectors at the same location but different images ($I_k$ and $I_{k-1}$) are highly correlated. This assumption is correct if the objects have a constant motion along several frames. In this case, in the first stage 16 vectors are clustered into 4 vectors (vectors 1 of image $I_{k-1}$, and 1' of $I_k$). In the next step, the four vectors (referred as 2) are clustered into one vector (named 3).

Tree construction is similar to the process for 2D clustering, although the vectors belong to different images. Other tree structures or clustering strategies can be applied in a similar way. This method can also be applied to other coding methods that work with image sequences. For instance, image sequence Block Truncation Coding.

## 5. CONCLUSIONS

We have used the quadtree structure in two different applications, with successful results. Basically, it is a method to adapt the block size without a significative increase of computational complexity.

The main advantages are:
- It is a simple and fast way to increase compression, breaking the compromise between block size and compression factor. Block size is chosen small for a good representation of detailed zones, and for the homogeneous zones the clustering with the quadtree implies greater block size.
- It is relatively easy to adapt this philosophy to different coders without modifying the structure of the basic coder. (It can be seen like a postprocessing algorithm over the basic compression method).

The main drawback of the clustering with quadtrees appears when there are transmission errors. In this case the undesired changes are visible every time the saved block is repeated. If the errors happen in the quadtree coding the result can be even worst.

## REFERENCES

[1] E. J. Delp & O. Robert Mitchell, "Image Compression Using Block Truncation Coding", IEEE Trans. Commun., Vol. COM-27, pp 1335-1342, September, 1979.

[2] B. Zeng, Y. Neuvo & A. N. Venetsanopoulos, "Interpolative BTC Image Coding", IEEE ICASSP, pp. 493-496, 1992.

[3] D. H. Healy & O. R. Mitchell, "Digital Video Bandwidth Compression Using Block Truncation Coding", IEEE Trans. Commun., Vol. COM-29, pp. 1809-1817, December, 1981.

[4] X. Domingo & M. Faúndez "Codificación Block Truncation con tamaño de bloque variable". URSI 95, pp.295-298.Valladolid. September 1995

[5] M. Faúndez, F. Vallverdú & F. Tarrés "Contributions to interframe coding". Workshop on image analysis and synthesis in Image coding. WIASIC 94 Berlin pp C3.1-C3.6

[6] M. Faúndez, F. Tarrés "A modification of the conjugate direction algorithm for motion estimation". EUSIPCO 92, pp.1311-1314.Brussels.